\documentstyle[11pt]{article}

\title{\bf The triple pomeron interaction\\ in the perturbative QCD}
\author{M.Braun \\ Department of high-energy
physics, \\ University of S. Petersburg, 198904 S. Petersburg
, Russia}
\def\beq{\begin{equation}}
\def\eeq{\end{equation}}
\def\noi{\noindent}

\begin{document}
\maketitle
\medskip
\noi{\bf Abstract.}

The triple pomeron interaction is studied in the perturbative approach of
BFKL-Bartels. At finite momentum transfers $\sqrt{-t}$ the contribution
factorizes
in the standard manner with a triple-pomeron vertex proportional to
$1/\sqrt{-t}$. At $t=0$ the contribution is finite, although it grows
faster with energy than for finite $t$ and does not factorize.

\vspace*{3 cm}
{\Large\bf SPbU-IP-1995/10}
\newpage
\section{Introduction}

There has been much theoretical activity around the triple pomeron
interaction, [ 1 - 4 ]
related to recent experimental data on the diffractive virtual photon
scattering on HERA [ 5 ]. In fact, the corresponding
diffractive differential cross-section involves many different
contributions, of which the triple pomeron  is only the
simplest one. All sorts of corrections can be added in the framework of
the standard Regge-Gribov theory, which include other pomerons or/and
$m$ to $n$ pomeron verteces. With the contribution of a single
pomeron exchange rising with energy, none of these unitarity corrections
can be safely considered as small. The situation improves in the
perturbative approach.
In the perturbative QCD one can study all contributions to
the diffractive cross-section ("effective triple pomeron vertex")
in the leading order in the small (fixed) coupling constant [ 2 ]. The
cross-section  turns out to be a rather complicated mixture of
the  triple pomeron interaction proper and  the transitions of four
interacting gluons into two pomerons, which makes its practical
calculation hardly possible.

Further improvement can, however, be achieved if along with a small
coupling constant one assumes a large number of colours
$N\rightarrow\infty$. Then, as one easily finds, the triple-pomeron
interaction indeed becomes the dominant contribution, all others damped
either by a factor $1/N^{2}$ (transitions 4 gluons $\rightarrow$ 2
pomerons) or by a supposedly small factor $g^{2}N$ ( $m$ to $n$ pomeron
verteces other than the triple one). One may hope that these assumptions
are not too far from reality. The triple pomeron interaction  thus
retains some practical interest, to say nothing of its theoretical
importance, as one of the basic quantities in the perturbative QCD at
high energies and small $x$.

Since in the BFKL theory [ 6 ] the pomeron is nonlocal and consists of
two interacting reggeized gluons, the triple pomeron vertex is
essentially a 2 to 4 gluon vertex in this theory. This vertex
$K_{2\rightarrow 4}$ was found by J.Bartels quite a time ago [ 7 ]
(and rederived for $t=0$ by a different method much later in [ 8 ]).
To obtain the triple pomeron vertex from $K_{2\rightarrow 4}$ all one
has to do is to integrate it with three BFKL pomerons coupled to
external sources and find the asymptotical behaviour of the resulting
expression at high energies where the pomerons exist. However this
straightforward (although requiring some care) program has not,
as far as we know,  been carried out up to now,  for reasons
beyond our comprehension. The only calculation of this sort in [ 8 ]
refers only to $t=0$ and is not explicit. The present note is devoted to
fill this gap.. We calculate the triple pomeron interaction in the
BFKL-Bartels approach, both at zero and nonzero momentum transfer
$\sqrt{-t}$. The result is finite in both cases although the point
$t=0$ is singular. For $t<0$ the contribution factorizes in the standard
manner into three pomerons coupled by the triple pomeron vertex
$\gamma_{3P}(t)=const/\sqrt{-t}$. The numerical constant which
determines the magnitude of $\gamma_{3P}$ is given by a complicated
6-dimensional momentum integral (Eq. (40 )), which we unfortunately have
not been able to calculate up to now. At $t=0$ the triple pomeron
contribution results much more complicated and does not factorize into
pomerons and their interaction vertex. It also rises with energy faster
than at $t<0$.

Comparing our results to the literature, we find much similarity to the
triple pomeron vertex at $t<0$ of A.Mueller an B.Patel [ 4 ], obtained
on the basis of the original equation for single and double colour
dipole densities proposed by A.Mueller in [ 9 ]. In particular the
$1/\sqrt{-t}$ behaviour is the same, which, however, does not mean much,
since this behaviour trivially follows from dimensional arguments. It is
the numerical factor that matters. This factor is different in [ 4 ]. As
we shall argue (see Section 5.) it corresponds to a different physical
picture, which, in our opinion, does not agree with the $s$-channel
unitarity.

\section{The triple pomeron interaction in the BFKL-Bartels formalism}

Consider the 3$\rightarrow$3 amplitude $A$ shown in Fig. 1. Let $s=(p_{1}-
p_{2})^{2}$, $s_{1}\equiv M^{2}=(p_{1}+p_{2}-p_{3})^{2}$, $l=p_{3}-p_{2}$,
$t=l^{2}$. We shall study the $s_{1}$ discontinuity of $A$
 in the triple Regge region $s,s_{1}\rightarrow\infty$,
$s>>s_{1}$, $t$ finite. It can be expressed via the unitarity relation
 in the $s_{1}$ channel
\beq
{\rm Disc}_{s_{1}}A\equiv 2iD=i\sum_{n}\int d\tau_{n} |A(3,n|1,2)|
^{2}
\eeq
The amplitude $A(3,n|1,2)$ refers to the process in which $n$ additional
gluons are produced with momenta $k_{1},...k_{n}$ (Fig. 2); $\tau_{n}$ is
the corresponding phase space
volume. In contrast to the analogous amplitude which appears in the
construction of the BFKL pomeron [ 6 ], in our case the lowest
$t$-channel is colourless. Our normalization is that the inclusive
diffractive cross-section to produce particle 3 with a missing mass
$M^{2}$ is given by
\beq
\frac{d\sigma}{dtdM^{2}}=\frac{D}{16\pi^{2}s^{2}}
\eeq

Such an amplitude with $n=1$ has been extensively studied by J.Bartels in
 [ 7 ] who has shown that in the leading order in the coupling constant
$g$  it can be found from its $s_{2}=(p_{3}+k)^{2}$ discontinuity.
 Suitably generalizing his result to $n$ produced gluons we assume that
 in the multiregge kinematical region the amplitude $A(3,n|1,2)$ can be
 represented
as shown in Fig. 3. Its upper part corresponds to a reggeized gluon which
emits intermediate real gluons. The lower part represents a pair of
interacting reggeized gluons in a colourless state, which form a pomeron.
The transitional vertex $V$ which describes emission of the gluon $k$, the
upper reggeon splitting into two lower ones, has been found in [ 7 ]. It
consists of two terms, a local and a nonlocal ones, both expressed through
the standard Lipatov reggeon-reggeon-particle vertex [ 6 ]. We shall not
need its explicit form here.

Putting Fig. 3 into the unitarity relation we find that the upper reggeized
gluon and its conjugated partner also form a pomeron, so that the
discontinuity $D$ can be graphically represented by Fig. 4. It shows three
pomerons coupled to colourless external sources and joined in the center
by a transitional vertex $K_{2\rightarrow 4}$. The latter is obtained by
squaring the Bartels vertex $V$ and summing over polarizations of the
intermediate gluon $k$. Explicitly, $K_{2\rightarrow 4}$, apart from a
colour factor and a factor $g^{4}$, is given by
\[
K_{2\rightarrow 4}(q_{1},-q_{1};q_{2},l-q_{2},-l+q_{3},-q_{3})\equiv
K_{l}(q_{1},q_{2},q_{3})=\]\beq
\frac{q_{1}^{2}q_{2}^{2}}{(q_{1}-q_{2})^{2}}+
\frac{q_{1}^{2}q_{3}^{2}}{(q_{1}-q_{3})^{2}}-
\frac{q_{1}^{4}(q_{2}-q_{3})^{2}}{(q_{1}-q_{2})^{2}(q_{1}-q_{3})^{2}}
\eeq
Translating Fig. 4 into an expression for $D$ we find
\[
D=(1/8)g^{10}N^{2}(N^{2}-1)(s^{2}/s_{1})
\int \prod_{i=1}^{3}(d^{2}q_{i}/(2\pi)^{3})K_{l}(q_{1},q_{2},q_{3})\]
\beq
\phi_{1}(s_{1},0,q_{1})\phi_{2}(s_{2},l,q_{2})\phi_{2}(s_{2},-l,-q_{3});
\ \ s_{2}=s/s_{1}
\eeq
Here $\phi_{1(2)}(s,l,q)$ is a pomeron (a solution to the BFKL equation)
 coupled
to the projectile (target) colourless external source with energy
squared  $s$,
total momentum $l$ and one of the gluon's momentum $q$. It is assumed
 that the colour factor for each source is $(1/2)\delta_{ab}$ and that
 each source
is proportional to $g^{2}$. The factor $g^{10}/8$ combines the
$g$-dependence from the sources  and the vertex $K$ and also
factors $1/2$ for each source from its colour structure.

Note that the BFKL equation determines $\phi(s,l,q)$ up to terms
proportional to $\delta^{2}(q)$ or $\delta^{2}(l-q)$. This circumstance is
of no importance when the pomeron is coupled to colourless sources at both
ends, since they vanish at $q=0$ or $q=l$. However, in our case this is
only true for gluons with momenta $q_{2}$ and $-q_{3}$: the vertex $K_{l}
(q_{1},q_{2},q_{3})$ is equal to zero when $q_{i}=0,\ i=1,2,3$ but it does
not vanish if $q_{2}=l$ or $q_{3}=l$. The physical solution, as obtained
from iterations of the BFKL equation, does not possess any singularities of
this type. So in the following we shall have to adopt special measures to
eliminate the $\delta$ type singularities at $q_{2}=l$ or $q_{3}=l$.

Since for $l\neq 0$ the solution of the BFKL equation is easier to obtain
 in the (transversal) coordinate space, we pass to this space by
presenting
\beq
\phi(s,l,q)=\int d^{2}r\phi(s,l,r)\exp (ir(q-l/2))
\eeq
Evidently $r$ is the transvercal distance between the gluons. Then (4)
transforms into
\[
D=(1/8)g^{10}N^{2}(N^{2}-1)(s^{2}/s_{1})(2\pi)^{-3}
\int \prod_{i=1}^{3}d^{2}r_{i}K_{l}(r_{1},r_{2},r_{3})\]
\beq\exp (-il(r_{2}+r_{3})/2)
\phi_{1}(s_{1},0,r_{1})\phi_{2}(s_{2},l,r_{2})\phi_{2}(s_{2},-l,-r_{3})
\eeq
where the 2 to 4 reggeon vertex in the coordinate space is obtained from
(3) to be
\[
K_{l}(r_{1},r_{2},r_{3})=
-2(\delta^{2}(r_{2}) \delta^{3}(r_{3})\nabla_{1}^{2}\delta^{2}(r_{1})+\]
\beq\delta^{2}(r_{3})r_{2}^{-2}(r_{2}\nabla_{1})\delta^{2}(r_{1})+
\delta^{2}(r_{2})r_{3}^{-2}(r_{3}\nabla_{1})\delta^{2}(r_{1})+
r_{2}^{-2}r_{3}^{-2}(r_{2}r_{3})\nabla_{1}^{4}\delta^{2}(r_{1}))
\eeq
The solutions $\phi$ vanish when $r=0$. So only the last term in (7)
survives.
We then can rewrite (6) as
\beq
D=(1/4)g^{10}N^{2}(N^{2}-1)(s^{2}/s_{1})
\int (d^{2}q/(2\pi)^{3})\chi_{1}(s_{1},0,q+l/2)
(\nabla_{q}\chi_{2}(s_{2},l,q))^{2}
\eeq
where
\beq
\chi_{1}(s,0,q)=\int d^{2}r\nabla^{4}\phi_{1}(s,o,r)\exp iqr
\eeq
and
\beq
\chi_{2}(s,l,q)=\int d^{2}r r^{-2}\phi_{2}(s,l,r)\exp iqr
\eeq

The solutions $\phi_{1(2)}$ can be obtained by using the Green function
of the BFKL equation for a given total momentum $G_{l}(s,r,r')$. For the
projectile (see [ 10 ]):
\beq
\phi_{1}(s,l,r)=\int(d^{2}r'/(2\pi)^{2})(G_{l}(s,r,0)-G_{l}(s,l,r')\
\rho_{1}(r')
\eeq
Here $\rho_{1}(r)$ is the colour density of the projectile as a function
of the intergluon distance with the colour factor $(1/2)\delta_{ab}$
and $g^{2}$ separated. For the projectile we choose a real or virtual
transverse photon with $p_{1}^{2}=-Q^{2}\leq 0$, which splits into
$q\bar q$ pairs of different flavours. The explicit form of $\rho$ is
well-known for this case [ 10 ]:
\beq
\rho_{1}(r)= (2/\pi)\alpha_{em}
\sum_{f=1}^{N_{f}}Z_{f}^{2}\int_{0}^{1}d\alpha
(m_{f}^{2}{\rm K}_{0}^{2}(\epsilon_{f} r)
+(\alpha^{2}+(1-\alpha)^{2})\epsilon_{f}^{2}{\rm K}_{1}^{2}
(\epsilon_{f} r))
\eeq
where $\epsilon_{f}^{2}=Q^{2}\alpha (1-\alpha)+m_{f}^{2}$ and $m_{f}$
and $Z_{f}$ are the mass and charge of the quark of flavour $f$.
For $Q^{2}>>m_{f}^{2}$ one can neglect the quark masses and evidently
\beq
\rho_{1}(r)=Q^{2}\tilde{f}_{1}(Qr)
\eeq
where $\tilde{f}_{1}(r)$ is a dimensionless function which behaves as
$1/r^{2}$ at the origin and exponentially falls at large $r$. For a real
photon ($Q=0$) $\rho_{1}$ has the same structure, with $q\rightarrow m$,
where $m$ is of the order of the lightest quark mass
\beq
\rho_{1}(r)=m^{2}f_{1}(mr)
\eeq
The dimensionless $f_{1}$ has the same properties as $\tilde{f}_{1}$.
As for the target, we shall not specify the explicit form of its colour
density $\rho_{2}(r)$. We shall only assume that the corresponding mass
scale is of the order $m$ and much less than $Q$ for a virtual
projectile, so that the scaling property (14) is valid for it.

Eqs (8)-(11) completely determine the triple pomeron interaction. In
what follows we calculate the asymptotical behaviour of $D$ at large
$s_{1}$ and $s_{2}$. To finally fix our normalizations we present
here the $2\rightarrow 2$ amplitude in the single pomeron exchange
approximation in our notations
\beq
A(s,t)=(i/4)g^{4}(N^{2}-1)s\int (d^{2}rd^{2}r'/(2\pi)^{4})\rho_{1}(r)
\rho_{2}(r')G_{l}(s,r,r')
\eeq
The elastic cross-section is given by
\beq
d\sigma^{el}/dt=(1/16\pi s^{2})|A(s,t)|^{2}
\eeq

\section{The triple pomeron vertex at $t<0$}

We begin with the calculation of the function $\chi_{1}(s,0,q)$ defined
by (9) and (11). Due to the singular character of $\nabla^{4}\phi$ as a
function of $r$ and the subtraction at $r'=0$ in (11) we shall need two
leading terms in the asymptotical expansion of $\chi_{1}$ at large $s$,
so that we have to actuate with some precision. The
BFKL Green function  at $l=0$ and large $s$ is given by the
expression [ 11 ]
\beq
G_{0}(s,r,r')=(1/8)rr'\int_{-\infty}^{\infty}\frac{d\nu s^{\omega(\nu)}}
{(\nu^{2}+1/4)^{2}}(r/r')^{-2i\nu}
\eeq
where
\beq
\omega(\nu)=(g^{2}N/2\pi^{2})(\psi(1)-{\rm Re}\psi(1/2+i\nu))
\eeq
Small values of $\nu$ play the dominant role in (17) at large $s$, so
that we can approximate
\beq
\omega(\nu)=\omega_{0}-a\nu^{2};\ \ \omega_{0}=(g^{2}N/\pi^{2})\ln 2,\ \
a=(7g^{2}N/2\pi^{2})\zeta (3)
\eeq
($\omega_{0}$ is the pomeron intercept minus one). Applying to (17) the
operator $\nabla^{4}$ we obtain
\beq
\nabla^{4}G_{0}(s,r,r')=(2s^{\omega_{0}}r'/r^{3})\int_{-\infty}^{\infty}
d\nu s^{-a\nu^{2}}(r/r')^{-2i\nu}
\eeq
To integrate over $r$ in (9) we shift the integration contour in $\nu$
from the real axis to a parallel line $C$ with ${\rm Im}\nu>1/2$. Then
(20) becomes integrable over $r$:
\beq
I_{1}\equiv\int d^{2}r\exp(iqr)\nabla^{4}G_{0}(s,r,r')=
\pi s^{\omega_{0}}qr'\int_{C}d\nu s^{-a\nu^{2}}(qr'/2)^{2i\nu}
\Gamma(-1/2-i\nu)/\Gamma (3/2+i\nu)
\eeq
Now we return to the integration over the real axis. In doing so we come
across a pole at $\nu=i/2$, which gives a separate contribution:
\beq
I_{1}=4\pi^{2}s^{\omega_{0}+a/4}+
\pi s^{\omega_{0}}qr'\int_{-\infty}^{\infty}
d\nu s^{-a\nu^{2}}(qr'/2)^{2i\nu}
\Gamma(-1/2-i\nu)/\Gamma (3/2+i\nu)
\eeq
The first term, coming from the pole, gives the leading contribution at
$s\rightarrow\infty$. It does not however depend on $r'$ and is
eliminated by the subtraction in (11). The second term is  standardly
evaluated by the stationary point method at large $s$:
\beq
-4\pi s^{\omega_{0}}\sqrt{\frac{\pi}{a\ln s}}
\exp (-\frac{\ln^{2}cqr'}{a\ln s})
\eeq
The number $c$ is determined from the equation for the stationary point
to be
\beq
c=2\exp(-2-\psi(1))
\eeq
Putting (23) into (11) we obtain
\beq
\chi_{1}(s,0,q)=4\pi qs^{\omega_{0}}
\sqrt{\frac{\pi}{a\ln s}}\int (d^{2}r/(2\pi)^{2})r\rho_{1}(r)
\exp (-\frac{\ln^{2}cqr}{a\ln s})
\eeq

Passing to the integration over $r$ in (25), let us first assume that
the projectile photon has a large virtuality $Q^{2}$. Using (13) we
change the integration over $r$ to that over the dimensionless $Qr$
\beq
\chi_{1}(s,0,q)=(4\pi qs^{\omega_{0}}/Q)
\sqrt{\frac{\pi}{a\ln s}}\int (d^{2}r/(2\pi)^{2})r\tilde{f}_{1}(r)
\exp (-\frac{\ln^{2}cqr/Q}{a\ln s})
\eeq
The integral over $r$ is now well convergent both at small and large
$r$. So $r$ takes on finite values. We shal later see that for $l\neq 0$
the integhral over $q$ is also convergent and average values of $q$ are
of order $l$. Therfore we may drop all factors in the logarithm squared
in the exponent, except $1/Q$. This finally gives
\beq
\chi_{1}(s,0,q)=4\pi qs^{\omega_{0}}(\tilde{b}_{1}/Q)
\sqrt{\frac{\pi}{a\ln s}}\exp (-\frac{\ln^{2}Q}{a\ln s})
\eeq
where the number $\tilde{b}_{1}$ is defined by
\beq
\tilde{b}_{1}=\int (d^{2}r/(2\pi)^{2})r\tilde{f}_{1}(r)
\eeq
Note that $R_{1}=\tilde{b}_{1}/Q$ has the meaning of the transverse
dimension of the projectile. From (12) we find
\[ \tilde{b}_{1}=(3\pi\alpha_{em}/256)\sum_{f}Z_{f}^{2}\]

The same argument evidently applies  to the case when the projectile
photon is real and $Q=0$. Then one uses (14) and obtains a similar
formula with $Q\rightarrow m$ and $\tilde{f}_{1}\rightarrow f_{1}$.
Since $m$ is finite one can also drop the exponential factor, so that
for a real photon
\beq
\chi_{1}(s,0,q)=4\pi qs^{\omega_{0}}(b_{1}/m)
\sqrt{\frac{\pi}{a\ln s}}
\eeq
where $b_{1}$ is defined as in (28) with
$\tilde{f}_{1}\rightarrow f_{1}$. For a real photon
\[ b_{1}=(3\alpha_{em}/64)m\sum_{f}Z_{f}^{2}/m_{f}\]

Now we turn to the function $\chi_{2}(s,l,q)$. The leading contribution
to the BFKL Green function at $l\neq 0$ has the form [ 11 ]
\beq
G_{l}(s,r,r')=(1/4\pi^{2})\int\frac{d\nu \nu^{2}}{(\nu^{2}+1/4)^{2}}
s^{\omega(\nu)}E_{l}^{\nu}(r)E_{l}^{-\nu}(r')
\eeq
where
\beq
E_{l}^{\nu}(r)=\int d^{2}R\exp(ilR)(\frac{r}{|R+r/2||R-r/2|})^{1+2i\nu}
\eeq
It is evident that the Green function (30), transformed into momentum
space, contains terms proportional to $\delta^{2}(l/2\pm q)$, which
should be absent in the physical solution (this circumstance was first
noted by A.H.Mueller and W.-K.Tang [ 12 ]). For that, (30) goes to zero
at $r=0$. Terms proportional to $\delta^{2}(l/2+q)$ are not dangerous
to us: they are killed by the vertex $K_{2\rightarrow 4}$, as noted in
the previous section. To remove the dangerous singularity at $q=l/2$
and simultaneously preserve good behaviour at $r=0$ we therefore make
a subtraction in $E$, changing it to
\beq
\tilde{E}^{\nu}_{l}(r)=
\int d^{2}R\exp(ilR)((\frac{r}{|R+r/2||R-r/2|})^{1+2i\nu}-
|R+r/2|^{-1-2i\nu}+|R-r/2|^{-1-2i\nu})
\eeq
This subtraction removes the $\delta$ singularity at $q=l/2$ and doubles
the one at $q=-l/2$, the latter eliminated by the kernel
$K_{2\rightarrow 4}$

Integration over $r$ leads to the integral
\beq
J(l,q)=\int (d^{2}r/(2\pi)^{2})(1/r^{2})\tilde{E}^{\nu}_{l}(r)
\eeq
This integral is convergent at any values of $\nu$, the point $\nu=0$
included, when the convergence at large values of $r$ and $R$ is
provided by the exponential factors. So in the limit
$s\rightarrow\infty$, when small values of $\nu$ dominate, we can put
$\nu=0$ in $J$:
\beq
J(l,q)=\int(d^{2}Rd^{2}r/(2\pi)^{2})\frac{\exp(ilR+iqr)}
{r|R+r/2||R-r/2|}+(1/l)\ln\frac{|l/2-q|}{|l/2+q|}
\eeq
The second term comes from the subtraction terms in (32). Passing to the
Fourier transform of the function $1/r$ one can represent (34) as an
integral in the momentum space
\beq
J(l,q)=(1/l)\int \frac{d^{2}p (l+|l/2+p|-|l/2-p|)
}{(2\pi)|l/2+p||l/2-p||q+p|}
\eeq

In the same manner we can put $\nu=0$ in the function $E^{-\nu}_{l}(r')$
obtaining for the integral over $r'$
\beq
\int \frac{d^{2}Rd^{2}r\exp(ilR)r\rho_{2}(r)}{(2\pi)^{2}|R+r/2||R-r/2|}=
(\pi/m)F(t)
\eeq
Here we have separated the characteristic dimensional factor $1/m$ (and
a factor $\pi$ for convenience) and introduced a dimensionless function
$F_{2}(t)$ which is a vertex for the interaction of the target with a
pomeron at momentum transfer $\sqrt{-t}$. $F_{2}(t)$ behaves like $\ln
t$ at small $t$. The rest of the Green function is easily calculated by
the stationary point method to finally give
\beq
\chi_{2}(s,l,q)=8s^{\omega_{0}}(\pi/a\ln s)^{3/2}(1/m)F_{2}(t)J(l,q)
\eeq

Combining our results for $\chi_{1}$ and $\chi_{2}$, from (8) we obtain
our final expression for the triple-pomeron interaction at $t<0$
\[
D=8\pi^{3/2}g^{10}N^{2}(N^{2}-1)(s^{2}/s_{1})
\frac{(s_{1}s_{2}^{2})^{\omega_{0}}}{sqrt{a\ln s_{1}}(a\ln s_{2})^{3}}
\]\beq
(\tilde{b}_{1}/Q)
\exp (-\frac{\ln^{2}Q}{a\ln s_{1}}
(F_{2}(t)/m)^{2}B/\sqrt{-t} \eeq
Here the number $B$ is defined as a result of the $q$ integration
\beq
B=l\int (d^{2}q/(2\pi)^{2})|l/2+q|(\nabla_{q}J(l,q))^{2}
\eeq
It does not depend on $l$ and can be represented as an integral over
three momenta
\beq
B=(1/(2\pi )^{4}l)\int d^{2}qd^{2}pd^{2}p'
\frac{|l/2+q|(q+p)(q+p')(l+p_{+}-p_{-})(l+p'_{+}-p'_{-})}
{p_{+}p_{-}p'_{+}p'_{-}|q+p|^{3}|q+p'|^{3}}
\eeq
where
 \beq p_{\pm}=|p\pm l/2|;\ \ p'_{\pm}=|p'\pm l/2|\eeq
It is a well-defined integral, although not quite easy to calculate.

To interprete various factors entering (38) we compare it with the
expressions for the elastic amplitude (15) both at $t=0$ and $t<0$ which
follow from similar calculations. At $t=0$, for a virtual photon
projectile,
\beq
A(s,0)=i(1/2)g^{4}(N^{2}-1)s^{1+\omega_{0}}(\tilde{b}_{1}b_{2}/mQ)
\sqrt{\frac{\pi}{a\ln s}}\exp (-\frac{\ln^{2}Q}{a\ln s})
\eeq
At $t<0$, for a real photon projectile
\beq
A(s,t)=i(1/2)g^{4}N^{2}s^{1+\omega_{0}}
\frac{\sqrt{\pi}}{(a\ln s)^{3/2}}(F_{1}(t)F_{2}(t)/m^{2})
\eeq
Evidently one cannot determine the pomeron and its couplings in a unique
way from these expressions. We may take that the pomeron contribution is
\beq P(s,t)=2\sqrt{\pi}s^{1+\omega_{0}}(a\ln s)^{-\epsilon (t)}\eeq
with $\epsilon (0)=1/2$ and $\epsilon (t)=3/2$ for $t<0$. Then taking
$N>>1$ we find
for its coupling $\gamma$ to a real external particle at $t=0$
\beq
\gamma(0)=(1/2)g^{2}Nb/m
\eeq
and at $t<0$
\beq
\gamma(t)=(1/2)g^{2}NF(t)/m
\eeq
For a coupling $\tilde{\gamma}$ to a virtual photon at $t=0$ we find
\beq
\tilde{\gamma}=(1/2)g^{2}N(\tilde{b}/Q)\exp (-\frac{\ln^{2}Q}{a\ln s})
\eeq
(it is not local in rapidity).

Rewriting (38) in terms of these quatities we obtain, say, for a virtual
projectile
\beq
D=\tilde{\gamma}_{1}\gamma_{2}^{2}(t)P(s_{1},0)P^{2}(s_{2},t)
\gamma_{3P}(t)
\eeq
with the triple-pomeron vertex $\gamma_{3P}(t)$ given by (for $N>>1$)
\beq
\gamma_{3P}(t)=8g^{4}NB/\sqrt{-t}
\eeq
Thus at $t<0$ the triple pomeron interaction factorizes in the standard
manner into three pomerons coupled to external sources and joined by
a triple pomeron vertex, independent of energies and proportional to
$1/\sqrt{-t}$, which dependence follows trivially from dimensional
considerations.

The triple pomeron interaction looks singular at $t=0$: both the vertex
$\gamma_{3P}$ and couplings $\gamma$ to external particles diverge at
$t=0$. However we shall presently see that the interaction is, in fact,
finite at $t=0$, although it grows with energy faster than (48).

\section{The triple pomeron interaction at $t=0$}

At $t=0$ we have to calculate the function $\chi_{2}$ in a different
manner, since the integrals (33) and (36) cease to converge. We take the
Green function (17) and integrate it over $r$, as indicated in (10)
\beq
I_{2}\equiv\int d^{2}r\exp (iqr)r^{-2}G_{0}(s,r,r')=
(\pi r'/4q)s^{\omega_{0}}\int \frac{d\nu s^{-a\nu^{2}}(qr')^{2i\nu}}
{(\nu^{2}+1/4)^{2}}\frac{\Gamma (1/2-i\nu)}{\Gamma(1/2+i\nu)}
\eeq
With $l=0$, $\nabla_{q} \chi_{2}=(q/|q|)(\partial/\partial q)\chi_{2}$.
So we actually need the derivative of (50) with respect to $q$.
Calculating the asymptotics at large $s$ we find
\beq
(\partial/\partial q)I_{2}=-(4\pi r'/q^{2})s^{\omega_{0}}
\sqrt{\frac{\pi}{a\ln s}}\exp (-\frac{\ln^{2}c_{1}qr'}{a\ln s})
\eeq
We do not need to shift the integration contour in $\nu$ here, since the
integral is convergent at small $r$. The number $c_{1}$, determined from
the stationary point equation, is
\beq
c_{1}=2\exp (-1-\psi(1))
\eeq
Integrating (51) over $r'$ with the external source we find
\beq
(\partial/\partial q)\chi_{2}(s,0,q)=
-(4\pi /q^{2})s^{\omega_{0}}
\sqrt{\frac{\pi}{a\ln s}}\int (d^{2}r/(2\pi)^{3})r\rho_{2}(r)
\exp (-\frac{\ln^{2}c_{1}qr}{a\ln s})
\eeq

Putting (25) and (53) into (8) we obtain the triple pomeron contribution
in the form of an integral over the projectile and target transverse
dimensions
\[
D(t=0)=2g^{10}N^{2}(N^{2}-1)(s^{2}/s_{1}) s_{1}^{\omega_{0}}
s_{2}^{2\omega_{0}}\sqrt{\frac{\pi}{a\ln s_{1}}}
\frac{\pi}{a\ln s_{2}}\]\beq
\int\prod_{i=1}^{3}(d^{2}r_{i}r_{i}/(2\pi)^{2})\rho_{1}(r_{1})
\rho_{2}(r_{2})\rho_{2}(r_{3})W(r_{1},r_{2},r_{3})
\eeq
where $W$ denotes the integral ober $q$:
\beq
W(r_{1},r_{2},r_{3})=\int (d^{2}q/(2\pi)^{2})q^{-3}
\exp (-\frac{\ln^{2}cqr_{1}}{a\ln s_{1}}
-\frac{\ln^{2}c_{1}qr_{2}}{a\ln s_{2}}-\frac{\ln^{2}c_{1}qr_{3}}{a\ln
s_{2}})
\eeq
It is easily calculated to give
\[
W=(1/2\pi)\sqrt{\pi
a_{0}}(r_{1}/e)^{\alpha_{1}}(r_{2}r_{3})^{\alpha_{2}}
\exp(a_{0}/4)\]\beq
\exp (\frac{a_{0}}{a_{1}a_{2}}(2\eta_{1}(\eta_{2}+\eta_{3}-
\eta_{1})-\eta_{2}^{2}-\eta_{3}^{2})-\frac{a_{0}}{a_{2}^{2}}
\ln^{2}(r_{2}/r_{3}))
\eeq
where we have denoted, for brevity,
\[\eta_{1}=\ln cr_{1},\ \eta_{2}=\ln c_{1}r_{2},\ \eta_{3}=\ln
c_{1}r_{3}\]
\[a_{1}=a\ln s_{1},\ a_{2}=a\ln s_{2},\ a_{0}=a\ln s_{1}\ln s_{2}/
\ln s_{1}s_{2}^{2}\]
\[\alpha_{1}=\ln s_{2}/\ln s_{1}s_{2}^{2},\ \alpha_{2}=
\ln s_{1}/\ln s_{1}s_{2}^{2},\ \alpha_{1}+2\alpha_{2}=1\]

{}From (56) we observe that at $t=0$ the triple pomeron interaction,
although finite, grows faster with energy than at $t<0$ due to the first
exponential factor. It leads to an additional power growth, whose
strength depends on the relation between $s_{1}$ and $s_{2}$.
 It is maximal
when $s_{1}\sim s_{2}\sim\sqrt{s}$ and this factor is $\sim s^{a/24}$.
One also notes that the dimensional factor is composed  from $r_{1},
r_{2}$ and $r_{3}$ in a proportion which also depends on the relation
between $s_{1}$ and $s_{2}$.

In the integration over $r$'s we, as before, assume that the only large
scale $Q$ may be involved in the virtual projectile. Then in the second
exponential in (56) we can drop all terms except $\eta_{1}$ which may be
substituted by $-\ln Q$. This leads to our final expression
\[
D(t=0)=(\pi/a) g^{10}N^{2}(N^{2}-1)(s^{2}/s_{1}) s_{1}^{\omega_{0}}
s_{2}^{2\omega_{0}}\sqrt{\frac{1}{\ln s_{2}\ln s_{1}s_{2}^{2}}}
(eQ)^{-1-\alpha_{1}}m^{-2(1+\alpha_{2})}\]\beq
\tilde{b}_{1}(\alpha_{1})b_{2}(\alpha_{2})
\exp(\frac{a\ln s_{1}\ln s_{2}}{4\ln s_{1}s_{2}^{2}}-
\frac{2\ln^{2}Q}{a\ln s_{1}s_{2}^{2}})
\eeq
where dimensionless $\tilde{b}_{1}(\alpha)$ and $b_{2}(\alpha)$ are
defined in analogy with (28) with an extra power of $r$:
\beq
\tilde{b}_{1}(\alpha)=\int(d^{2}r/(2\pi)^{3})r^{1+\alpha}\tilde{f}_{1}(r)
\eeq
and similarly for $b_{2}(\alpha)$.

The expression (57) has a different structure as compared with (47) at
$t<0$. It does not factorize into three pomerons and their interaction
vertex independent of energies. In fact the dependence of (57) on
$s_{1}$ and $s_{2}$ is quite complicated and also enters the target and
projectile factors $b$. However this loss of factorization should only
occur at very small values of $\sqrt{-t}$. In the integral (55) very
small values of $q$ dominate: $\ln (1/q)\sim\sqrt{\ln s}$ ,i.e.
$q\sim m\exp(-\sqrt{a\ln s})$. Recalling that at $t<0$ characteristic
values of $q$ are of order $l$, we come to the conclusion that the
transition from (47) to (57) and the breakdown of factorization occurs
at
\beq
\sqrt{-t}< m\exp(-\sqrt{a\ln s})
\eeq

\section{Discussion}

We have calculated the triple pomeron interaction in the perturbative
QCD picture, based on the gluon reggeization and $s$-channel unitarity
(the BFKL-Bartels approach). The important result is that, contrary to
some pessimistic judgements [ 13 ], the triple pomeron vertex is
infrared finite both at $t=0$ and $t<0$. The subtraction (32) in the
pomeron Green function has been essential in this respect.

It is instructive to compare our result (at $t<0$) to that of
A.H.Mueller and B.Patel [ 4 ], which was obtained in a different
framework, based on original evolution equations for single and
double dipole densities [ 9 ]. We shall not attempt to discuss these
equations in general here, limiting ourselves with only the triple
pomeron, as obtained in [ 4 ]. At $t<0$ the triple pomeron interaction
of A.H.Mueller and B.Patel is the same as our Eq. (38) (for a real
projectile) with a different numerical factor:
\beq
B\rightarrow (1/2\pi^{2})V_{0}
\eeq
where, in our notation,
\beq
V_{0}=(l/\pi)\int (d^{2}r_{2}d^{2}r_{3}/(2\pi)^{2})
\exp (-il(r_{2}+r_{3})/2)
\frac{E_{l}^{0}(r_{2})E_{l}^{0}(r_{3})}{|r_{2}+r_{3}|r_{2}^{2}r_{3}^{2}}
\eeq
Turning to our derivation, it is easy to check that this result is
obtained if, instead of the Bartels vertex (7), one uses
\[
\tilde{K}_{l}(r_{1},r_{2},r_{3})=-(\pi/4)
\frac{\delta^{2}(r_{1}+r_{2}+r_{3})}{r_{1}^{2}r_{2}^{2}r_{3}^{2}}+\]
\beq terms\ proportional\ to\ \delta^{2}(r_{i}),\ i=1,2,3
\eeq
This simple vertex has the form which one might naively write from
dimensional arguments, assuming also symmetry in the interacting
pomerons. It is, however, singular in the ultraviolet and does not admit
transition to the momentum space. It is evidently different from the
Bartels vertex (7) obtained from the $s$-channel unitarity. Thus it
looks as if the triple pomeron of A.H.Mueller and B.Patel would not
correspond to the $s$-channel unitarity.

 \section
{Acknowledgements}

The author is grateful to Misha Ryskin for valuable
discussions.

 \newpage

\section{References}

1. E.Levin and M.Wuesthoff, preprint DESY 92-166 (1992).\\
2. J.Bartels and M.Wuesthoff, preprint DESY 94-016 (1994).\\
3. N.N.Nikolaev and B.G.Zakharov, preprint KFA-IKP(Th)-1993-17 (1993).\\
M.Genovese, N.N.Nikolaev and B.G.Zakharov, preprint CERN-TH/95-12
 (1995).\\
4. A.H.Mueller and B.Patel, Nucl. Phys., {\bf B425} (1994) 471.\\
5. ZEUS collaboration: M.Derrick et al., Phys. Lett. {\bf B315} (1993)
481; H1 collaboration: T.Ahmed et al., Nucl. Phys. {\bf B429} (1994)
477.\\
6. V.S.Fadin, E.A.Kuraev and L.N.Lipatov, Phys. Lett. {\bf B60} (1975)
50.\\
I.I.Balitsky and L.N.Lipatov, Sov.J.Nucl.Phys. {\bf 15} (1978) 438.\\
7. J.Bartels, Nucl. Phys. {\bf B175} (1980) 365.\\
8. E.M.Levin and M.G.Ryskin, Phys. Rep. {\bf 189} (1990) 267-382 (App.
C).\\
9. A.H. Mueller, Nucl. Phys. {\bf B415} (1994) 373.\\
10. N.N.Nikolaev and B.G.Zakharov, Z. Phys. {\bf C 49} (1991) 607.\\
11. L.N.Lipatov, Zh.Eksp.Teor.Fiz. {\bf 90} (1986) 1536 (Sov. Phys.
JETP {\bf 63} (1986) 904).\\
12. A.H.Mueller and W.K.-Tang, Phys. Lett. {\bf B 284} (1992) 123.\\
13. L.N.Lipatov, in: Perturbative QCD, ed. A.H.Mueller
(World Scientific, Singapore, 1989).\\

\newpage
\section{Figure captions}

\noi Fig. 1. The $3\rightarrow 3$ amplitude whose discontinuity in
$M^{2}=(p_{1}+p_{2}-p_{3})^{2}$  contains the triple pomeron
interaction.\\
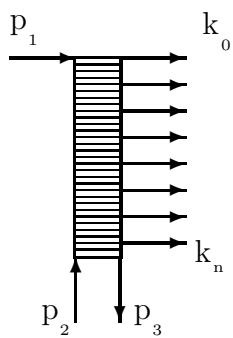
Fig. 2. The production amplitude $A(3,n|12)$ which enters the unitarity
equation (1).\\
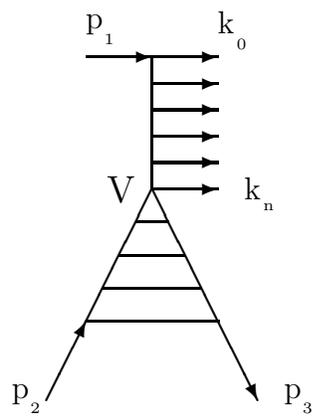
Fig. 3. The production amplitude $A(3,n|1,2)$ in the multiregge
kinematical region. The upper part shows a reggeized gluon emitting real
gluons. The lower part shows two interacting reggeized gluons, which
form a pomeron. The two parts are joined by a transitional vertex $V$.\\
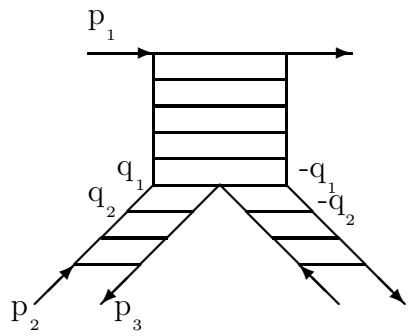
Fig. 4. The triple pomeron interaction as a result of a transition from
two reggeized gluons to four. The central line, with the
emission of two additional gluons from a point, corresponds to the
Bartels vertex $K_{2\rightarrow 4}$, Eq. (3).\\

 \newpage


\begin{picture}(200,200)(-10,-10)
\thicklines
\put(50,50){\framebox(50,75)}
\put(25,125){\vector(1,0){25}}
\put(100,125){\vector(1,0){25}}
\put(50,25){\vector(0,1){25}}
\put(66.7,50){\vector(0,-1){25}}
\put(83.4,25){\vector(0,1){25}}
\put(100,50){\vector(0,-1){25}}
\put(66.7,0){\Large Fig. 1}
\thinlines
\multiput(50,52.5)(0,2.5){29}{\line(1,0){50}}
\put(25,130){\begin{picture}(15,15)
\put(0,6){{\large p }}
\put(7,0){{\tiny  1}}
\end{picture}}
\put(125,130){\begin{picture}(15,15)
\put(0,6){{\large p }}
\put(7,0){{\tiny  1}}
\end{picture}}
\put(37,20){\begin{picture}(15,15)
\put(0,6){{\large p }}
\put(7,0){{\tiny  2}}
\end{picture}}
\put(56.7,20){\begin{picture}(15,15)
\put(0,6){{\large p }}
\put(7,0){{\tiny  3}}
\end{picture}}
\put(73.7,20){\begin{picture}(15,15)
\put(0,6){{\large p }}
\put(7,0){{\tiny  3}}
\end{picture}}
\put(105,20){\begin{picture}(15,15)
\put(0,6){{\large p }}
\put(7,0){{\tiny  2}}
\end{picture}}
\end{picture}


\begin{picture}(200,200)(-10,-10)
\thicklines
\put(50,50){\framebox(16.7,75)}
\put(25,125){\vector(1,0){25}}
\multiput(66.7,125)(0,-10){8}{\vector(1,0){25}}
\put(50,25){\vector(0,1){25}}
\put(66.7,50){\vector(0,-1){25}}
\put(60,0){\Large Fig. 2}
\thinlines
\multiput(50,52.5)(0,2.5){29}{\line(1,0){16.7}}
\put(25,130){\begin{picture}(15,15)
\put(0,6){{\large p }}
\put(7,0){{\tiny  1}}
\end{picture}}
\put(98,128){\begin{picture}(15,15)
\put(0,6){{\large k }}
\put(7,0){{\tiny  0}}
\end{picture}}
\put(37,20){\begin{picture}(15,15)
\put(0,6){{\large p }}
\put(7,0){{\tiny  2}}
\end{picture}}
\put(72,20){\begin{picture}(15,15)
\put(0,6){{\large p }}
\put(7,0){{\tiny  3}}
\end{picture}}
\put(95,42){\begin{picture}(15,15)
\put(0,6){{\large k }}
\put(7,2){{\tiny  n}}
\end{picture}}
\end{picture}


\begin{picture}(200,200)(-10,-10)
\thicklines
\put(100,100){\line(0,1){50}}
\put(75,150){\vector(1,0){25}}
\multiput(100,150)(0,-10){6}{\vector(1,0){25}}
\put(100,100){\line(-1,-2){25}}
\put(93.75,87.75){\line(1,0){12.5}}
\put(87.50,75){\line(1,0){25}}
\put(81.25,62.5){\line(1,0){37.5}}
\put(75.50,50){\line(1,0){50}}
\put(100,100){\line(1,-2){25}}
\put(60,20){\vector(1,2){15}}
\put(125,50){\vector(1,-2){15}}
\put(95,0){\Large Fig. 3.}
\put(83,95){\Large V}
\put(75,155){\begin{picture}(15,15)
\put(0,6){{\large p }}
\put(7,0){{\tiny  1}}
\end{picture}}
\put(125,153){\begin{picture}(15,15)
\put(0,6){{\large k }}
\put(7,0){{\tiny  0}}
\end{picture}}
\put(47,15){\begin{picture}(15,15)
\put(0,6){{\large p }}
\put(7,0){{\tiny  2}}
\end{picture}}
\put(150,15){\begin{picture}(15,15)
\put(0,6){{\large p }}
\put(7,0){{\tiny  3}}
\end{picture}}
\put(135,90){\begin{picture}(15,15)
\put(0,6){{\large k }}
\put(7,2){{\tiny  n}}
\end{picture}}
\end{picture}


\begin{picture}(200,200)(-10,-10)
\thicklines
\put(50,100){\line(0,1){50}}
\put(100,100){\line(0,1){50}}
\put(25,150){\vector(1,0){25}}
\put(100,150){\vector(1,0){25}}
\multiput(50,150)(0,-10){6}{\line(1,0){50}}
\put(50,100){\line(-1,-1){30}}
\put(75,100){\line(-1,-1){30}}
\multiput(40,90)(-10,-10){3}{\line(1,0){25}}
\put(100,100){\line(1,-1){30}}
\put(75,100){\line(1,-1){30}}
\multiput(85,90)(10,-10){3}{\line(1,0){25}}
\put(5,55){\vector(1,1){15}}
\put(45,70){\vector(-1,-1){15}}
\put(120,55){\vector(-1,1){15}}
\put(130,70){\vector(1,-1){15}}
\put(70,20){\Large Fig. 4}
\put(25,155){\begin{picture}(15,15)
\put(0,6){{\large p }}
\put(7,0){{\tiny  1}}
\end{picture}}
\put(36,98){\begin{picture}(15,15)
\put(0,6){{\large q }}
\put(7,0){{\tiny  1}}
\end{picture}}
\put(25,88){\begin{picture}(15,15)
\put(0,6){{\large q }}
\put(7,0){{\tiny  2}}
\end{picture}}
\put(-4,45){\begin{picture}(15,15)
\put(0,6){{\large p }}
\put(7,0){{\tiny  2}}
\end{picture}}
\put(35,45){\begin{picture}(15,15)
\put(0,6){{\large p }}
\put(7,0){{\tiny  3}}
\end{picture}}
\put(105,97){\begin{picture}(20,20)
\put(0,6){{\large -q }}
\put(10.5,0){{\tiny  1}}
\end{picture}}
\put(112,85){\begin{picture}(20,20)
\put(0,6){{\large -q }}
\put(10.5,0){{\tiny  2}}
\end{picture}}
\end{picture}

 
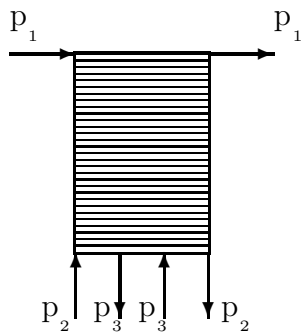
\end{document}